\documentstyle[epsfig]{aipproc}

\begin{document}
\title{Magnetic Stresses at the Inner Edges of Accretion Disks Around Black
Holes}

\author{Julian H. Krolik$^*$}
\address{$^*$Department of Physics and Astronomy, Johns Hopkins University,
Baltimore MD 21209}

\maketitle

\begin{abstract}

     For the past twenty-five years, nearly all analyses of accretion disk
dynamics have assumed that stress inside the disk is locally proportional
to pressure (the ``$\alpha$-model") and that this stress goes to zero
at the marginally stable orbit.  Recently, it has been demonstrated
that MHD turbulence accounts for the bulk of internal disk stress.
In contradiction with the traditional view, the stress from this
MHD turbulence does not diminish near the marginally stable orbit,
and the ratio of magnetic stress to pressure rises sharply there.   Examples of
the consequences include: an increase in accretion efficiency that
may also be time- and circumstance-dependent; a decrease in the
rate of black hole spin-up by accretion; and generation of disk
luminosity fluctuations.  Preliminary results from numerical simulations
lend support to analytic estimates of these effects.

\end{abstract}

\section{Introduction}

    Ever since the seminal work of Novikov and Thorne \cite{jhk:nt73} and
Shakura and Sunyaev \cite{jhk:ss73}, our understanding of the physics of
accretion disks around black holes has been built on two fundamental beliefs: the
energy available for radiation is equal to the binding energy of the
innermost stable orbit, and is therefore determined solely by the
spin of the black hole; and the (vertically-integrated) stress responsible
for transporting angular momentum outwards through the disk is proportional
to the local pressure.  Recent work involving magnetic forces in the
innermost regions of relativistic accretion disks has undercut both of
these claims.  In this review I will summarize the reasons for this
change of heart and explain the consequences that are likely to follow
if the new point of view proves correct.

\section{The Traditional View of the Inner Edges of Relativistic Accretion
Disks}

    The basic framework for understanding the inner regions of relativistic
accretion disks was laid out more than twenty-five years ago \cite{jhk:nt73,%
jhk:pt74}.  In the simplest possible picture (i.e., one assuming that disks
are time-steady, axi-symmetric, and geometrically thin), their radial
structure may be defined in terms of conservation equations that appear
almost Newtonian, as the general relativistic effects can be collapsed into
multiplicative correction factors.  Conservation of angular momentum,
for example, may be expressed through a single first-order differential
equation in radius.

    Thus, to specify the entire solution, all that is necessary is to choose
a boundary condition for this differential equation.  This boundary condition
may be physically interpreted as determining the conserved outward angular
momentum flux through the disk, or, equivalently, as specifying $T_{r\phi}$,
the $r$--$\phi$ component of the stress exerted on the disk, at its inner edge.
This inner edge is conventionally taken to lie at the radius of the marginally
stable orbit, $r_{ms}$.

   Selecting this boundary condition is the only part of the procedure
in which there is any guesswork.  Until very recently, it was almost universally
assumed that the stress must be zero at $r_{ms}$.  Two reasons were commonly
cited for this choice.
The first, due to Thorne and collaborators \cite{jhk:nt73,jhk:pt74},
was that the radial speed inside $r_{ms}$ is so much larger than the
radial speed outside $r_{ms}$ that, assuming
a constant mass accretion rate, the inertia of mass inside $r_{ms}$ must be much
smaller than the inertia of the disk proper in the region of stable orbits.
Given that, it appears difficult to see how the small amount of matter in
the plunging region could exert any significant force on the much heavier
disk.  The second argument \cite{jhk:ak89} begins with the assumption
that the stress carrying angular momentum outward should always be
proportional to the local pressure; i.e., $T_{r\phi} = \alpha p$\cite{jhk:ss73}.
If this is so, as material accelerates inward and expands, its pressure
falls rapidly, so the stress must do likewise.

    If there is no stress inside $r_{ms}$, then no forces
change the energy or angular momentum of material inside that point.
Matter therefore arrives at the event horizon with the energy and angular momentum
it had at $r_{ms}$.  The maximum energy per unit mass available to be radiated
in the disk (the maximum radiative efficiency) is then defined simply by
the binding energy of the marginally stable orbit, and the spin-up rate
of the black hole is the mass accretion rate times the specific angular
momentum of that orbit.  Both the specific binding energy and the
specific angular momentum at $r_{ms}$ depend only on the black hole's
normalized angular momentum $a/M$; thus, it has long been thought that both
the radiative efficiency and the spin-up rate per unit accreted mass
are functions only of $a/M$. 

   Interestingly, the very first work on this subject \cite{jhk:pt74} acknowledged
that strong magnetic fields could upset the arguments that the stress must
go to zero at $r_{ms}$,
but chose (reasonably, given the state of knowledge at the time) to ignore this
possibility.  However, work of the past decade (summarized in \cite{jhk:bh98})
has shown that magnetic fields are essential to angular momentum transfer
in accretion disks.  We must therefore re-open the question of the
appropriate boundary condition on $T_{r\phi}$ at $r_{ms}$.

\section{Magnetic Fields in Accretion Disks}

    This recent work has shown that, in the main body of the disk, MHD
turbulence creates a stress that, when vertically-averaged, is
roughly proportional to the pressure, with an
$\alpha \sim 0.01$ -- 0.1 \cite{jhk:bh98,jhk:s96,jhk:b96}.  The magnetic stress
accounts for the bulk of the angular momentum flux so long as two
conditions are met: there must be some seed magnetic field in the accreting
gas, but in the disk midplane the energy density of any imposed
magnetic field must be smaller than the total pressure; and the matter must
have high enough conductivity that the MHD approximation is valid.  In the
conditions surrounding black holes, these are all almost certain to apply.
Weak magnetic fields are virtually ubiquitous in astrophysical environments;
and near a black hole the gas temperature will very nearly always be high enough
to maintain the gas in an ionized, and therefore highly conducting, state.

    This turbulence is generated by an MHD instability driven by the orbital
shear \cite{jhk:bh91}.  Its growth rate is always comparable to the
orbital frequency, so it grows roughly as rapidly (relative to an orbital
period) in the relativistic part of the disk as in the Newtonian part.
The nonlinear amplitude of the turbulence in the Newtonian part of the
disk is determined by a cascade of energy toward shorter wavelengths,
where it can ultimately be dissipated.  This process, too, should operate more
or less in the same fashion (as viewed in the fluid frame) in the
relativistic part of the disk as in the non-relativistic part, provided
the local inflow time is long enough for equilibrium to be established.
We may therefore reasonably conclude that the ratio $B^2/(8\pi p)$
doesn't change dramatically at radii a few times $r_{ms}$.
Because orbital shear automatically stretches field lines in the sense
that produces outward angular momentum flow ($\langle B_r B_\phi \rangle < 0$),
the effective $\alpha$-parameter ($= \langle B_r B_\phi \rangle/(4\pi p)$)
is also unlikely to change much.  But if this is so, {\it why should
the stress diminish as the marginally stable orbit is approached?}

    The only aspect of this process that changes in any qualitative way
as $r_{ms}$ is approached from the outside is the ordering of four timescales:
the inflow time $t_{in}$, the thermal time $t_{th}$, the turbulent dissipation
time $t_{diss}$, and the dynamical time $t_{dyn}$.  In the disk body,
$t_{in} \sim \alpha^{-1} (r/h)^2 t_{dyn}$ and $\beta t_{diss} \sim t_{th} \sim
\alpha^{-1} t_{dyn}$, where $\beta$ is the ratio of gas (+ radiation)
pressure to magnetic
energy density.  Thus, $t_{in} \gg t_{th} > t_{diss} > t_{dyn}$.
However, just outside $r_{ms}$
the effective potential flattens (this is, of course, what it means for
$r_{ms}$ to be the location of the innermost {\it marginally stable}
orbit), so $t_{in}$ diminishes toward $t_{dyn}$.  The concomitant
decline in surface density likewise reduces $t_{th}$.  However,
$t_{diss}$ (crudely $\sim h/v_A$ for disk thickness $h$ and Alfven
speed $v_A$) declines more slowly than $t_{th}$. As a result, in the
vicinity of $r_{ms}$, the interplay of plasma dynamics and flux-freezing
should be at least as important to determining the field intensity as the
balance between the turbulent dynamo and turbulent dissipation.

     If flux-freezing really does determine the evolution of the field
as the matter plunges inside $r_{ms}$, the magnetic field strength
in the fluid frame stays roughly constant or increases somewhat even
while the fluid pressure decreases dramatically \cite{jhk:k99}.  It
follows that the effective $\alpha$ rises sharply as the inflow
accelerates near and inside $r_{ms}$.  In fact,
when the radial component of the velocity becomes relativistic, it
is easy to show that these assumptions imply
$B^2/(8\pi) \sim \rho c^2$.  That is, when the inflow is relativistic,
magnetic forces become competitive
with gravity, and the Alfven speed becomes relativistic.  Matter may then
retain a causal coupling with the disk it left behind even when it has
fallen well inside $r_{ms}$, allowing significant transfer of energy and angular
momentum \cite{jhk:k99,jhk:g99}.

    Looking back at the old argument that the small inertia of mattter in the
plunging region cannot do much to the main body of the disk, we now see that
magnetic fields in effect turn this argument on its head.  Magnetic
connections between the disk and plunging matter act as ``tethers" by which
the massive disk restrains angular acceleration of the low-inertia streams
inside $r_{ms}$.  In so doing, stress can be exerted on the
disk itself.

\section{Consequences}

    The existence of significant stress at the marginally stable orbit has
major consequences for the most fundamental properties of accretion onto black
holes.  Energy taken from plunging matter and delivered to the disk is
potentially available for radiation; removal of angular momentum from this
matter retards the rate of black hole spin-up \cite{jhk:ak00}.  In fact, when the
black hole is rapidly rotating, the rotational energy of the black hole
itself can be tapped: frame-dragging gives plunging matter a high
orbital frequency; magnetic connections from this region to the disk
exert a torque; the end-result is energy drawn from black hole rotation given
to matter in the disk \cite{jhk:g99,jhk:k00}.  In other words, the amount
of energy drawn from accreting matter is no longer a quantity fixed by
orbital mechanics: it is a dynamical quantity that is the product of
complicated MHD dynamics, and may even be time-variable.

    The fate of the work done by magnetic fields at $r \geq r_{ms}$ depends
critically on the ratio between the dissipation and inflow timescales in the
place where the energy is brought.  If the energy can be converted into heat
and radiated before the matter finds its way into the plunging region inside
$r_{ms}$, it adds to the radiative efficiency; if, on the other hand, dissipation
is slow, so that the energy stays in the gas either as organized kinetic
energy or as magnetic field energy, it may end up being taken into the
black hole.  Given the arguments of the previous section, it is as yet
unclear how this balance works out.
    
    If additional heat is deposited in the disk, there can be substantial
alterations to the disk spectrum \cite{jhk:ak00}.  Additional flux is
radiated at high frequencies, and relativistic effects beam it strongly into
the equatorial plane.  In addition, because so much more radiation is produced
in the most relativistic portion of the disk, the returning radiation fraction
is greatly enhanced.  This latter effect has implications for phenomenology
as diverse as polarization of the emitted light and the synchronization of
fluctuations.

     However, there is an additional implication of stress at the inner edge
of the disk that may be important even if the radiative efficiency is hardly
altered by these mechanisms: dynamics at the marginally stable orbit can
be a powerful ``noise" source to the entire dynamical system.  In the long
run, these dynamical fluctuations may be reflected in luminosity
fluctuations, which are, of course, a hallmark of all known accreting
black holes.  The origin of this ``noise" may be seen from a simple thought
experiment: Consider a small magnetized fluid element orbiting just
outside $r_{ms}$.  Imagine that half this fluid element receives a negative
angular momentum perturbation from the turbulence and begins to fall inward.
As it does so, its orbital frequency increases simply due to its decreasing
orbital radius.  The resulting shear between the two halves of the fluid
element stretches the magnetic flux tube connecting them, and a magnetic
tension force creates a torque that transfers angular momentum from the
falling half to the half that is still orbiting stably.  The stably orbiting
half must then move outward, launching a compressive wave into the disk.
Thus, the basic mechanics of accretion create disk ``noise" when there are
magnetic connections across the marginally stable orbit.

\section{Simulations}

    All the arguments presented so far have been qualitative.  Whether
these effects are quantitatively important can only be ascertained as a
result of genuine calculation.  MHD turbulent dynamics are sufficiently
complicated that (almost) all credible calculations are
numerical simulations.  Although much work remains
to be done before simulations can be run with a requisite level of
realism, some preliminary results have been achieved
\cite{jhk:h00,jhk:hk01,jhk:arc01}.

     To date, all simulations touching on these issues have been limited
in a number of regards, both physical and numerical.  All have
assumed Newtonian physics; general relativistic dynamics is mimicked
by the use of the Paczy\'nski-Wiita potential $U = -GM/(r - 2GM/c^2)$.
This potential qualitatively reproduces motion in a Schwarzschild
metric by creating a marginally stable orbit at $r=6GM/c^2$.  In
addition, all simulations so far have substituted an assumed equation
of state for a real energy equation.  This assumption has two deleterious
consequences: magnetosonic waves don't propagate at the right speed
because the pressure isn't correctly computed; and, more importantly,
it is impossible to use these simulations to estimate the radiative
efficiency because energy taken from the plunging region to the disk
falls right back in.  There are also aspects of the numerical
methods used in these simulations that are less than optimal.
Both kinetic energy and magnetic energy can disappear if fluid from
two adjacent cells with oppositely directed velocity or magnetic field
components is combined.  It also appears likely that no simulation
to date has used a grid fine enough to resolve the magnetic
field structure.

\begin{figure}[t]
\centerline{\epsfig{file=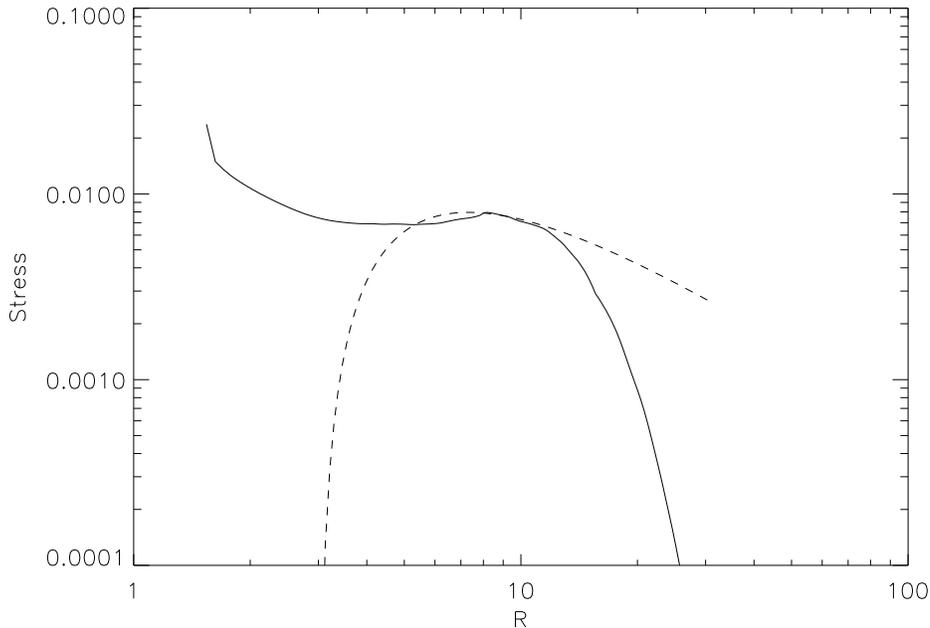,angle=90,width=5in}}
\caption{The solid curve shows the time- and azimuthal-average of the
vertically-integrated $r$-$\phi$ component of the magnetic stress in
a simulation by Hawley and Krolik \protect\cite{jhk:hk01}.  The dashed curve is
the prediction of the Novikov-Thorne model for the mean mass accretion
rate found in that simulation.  Note that the unit of distance is
$2GM/c^2$, so the marginally stable orbit is at $R=3$.}
\label{jhk:avestress}
\end{figure}

    Nonetheless, putting aside these qualms, the simulational results may
still be
used as an indication of what may appear in more realistic simulations.
In particular, angular momentum transport should be more reliably
treated than energy transport because it is not subject to the problems
described in the previous paragraph.  The result shown in
Figure~\ref{jhk:avestress} \cite{jhk:hk01} is particularly noteworthy.
Two distributions
of $T_{r\phi}$ integrated vertically and averaged over both azimuth and
time are shown: the prediction of the Novikov-Thorne model for the mean
mass accretion rate found in the simulation; and the magnetic contribution
to the stress.  The two curves coincide very closely over the radial range
from $\simeq 8$ -- $24 GM/c^2$, demonstrating that magnetic
stresses account for very nearly all the angular momentum transport
in the main part of the disk.  They depart in an uninteresting way at
large radius---the deviation here is due to the fact that the disk in
the simulation had an outer edge, whereas the Novikov-Thorne model refers
to a disk that extends to very large radius.   At small radius, however,
the contrast is very significant---as expected on the basis of the qualitative
arguments presented earlier, magnetic stress remains important across
the marginally stable orbit and throughout the region where gas plunges
toward the event horizon.  Indeed, as a result of this continuing stress,
the mean specific angular momentum of matter crossing the inner edge of the
simulation (at $r = 3GM/c^2$) is about 5\% smaller than the specific
angular momentum at $r_{ms}$.  Also as expected, the effective $\alpha$
parameter increases sharply in the innermost part of the accretion flow:
between $r=8GM/c^2$ and $r=4GM/c^2$, it increases by an order of magnitude.

\section{Future Prospects}

    The results of these first simulations are encouraging, but they are
far from the final answer.  Fortunately, significant improvements
are feasible.  Improved resolution can be obtained both by cleverer
gridding schemes and, of course, by a few years' technological
development.  Real energy equations can be computed by explicitly
incorporating phenemonological viscosity and resistivity and a
radiative cooling function (following time-dependent radiation transfer
is also possible \cite{jhk:ts01}, but will require somewhat greater
computer power before it is feasible for this kind of simulation).
Likewise, there is no fundamental impediment to writing MHD codes that
work in a fixed relativistic metric.  Thus, in a few years, we should
be able to attach quantitative values to the effects pointed out qualitatively
here.

\end{document}